\documentclass{article}

\usepackage{arxiv}

\usepackage[utf8]{inputenc} % allow utf-8 input
\usepackage[T1]{fontenc}    % use 8-bit T1 fonts
\usepackage{hyperref}       % hyperlinks
\usepackage{url}            % simple URL typesetting
\usepackage{booktabs}       % professional-quality tables
\usepackage{amsfonts}       % blackboard math symbols
\usepackage{microtype}      % microtypography
\usepackage{graphicx}
\usepackage{float}
\usepackage{amsmath}
\title{WideDTA: prediction of drug-target  binding affinity}

\author{
  Hakime ~\"{O}zt\"{u}rk  \\
  Department of Computer Engineering\\
  Bogazici University\\
  34342 Istanbul, Turkey \\
  \texttt{hakime.ozturk@boun.edu.tr} \\
   \And
 Elif ~Ozkirimli \\
  Department of Chemical Engineering\\
  Bogazici University\\
  34342 Istanbul, Turkey \\
  \texttt{elif.ozkirimli@boun.edu.tr} \\
     \And
 Arzucan ~\"{O}zg\"{u}r \\
  Department of Computer Engineering\\
  Bogazici University\\
  34342 Istanbul, Turkey \\
  \texttt{arzucan.ozgur@boun.edu.tr} \\
}

\begin{document}
\maketitle

\begin{abstract}
\textbf{Motivation:}  Prediction of the interaction affinity between proteins and compounds is a major challenge in the drug discovery process.  WideDTA is a deep-learning based prediction model that employs chemical and biological textual sequence information to predict binding affinity. \\
\textbf{Results:}  WideDTA uses four text-based information sources, namely the protein sequence, ligand SMILES, protein domains and motifs, and maximum common substructure words to predict binding affinity. WideDTA outperformed one of the state of the art deep learning methods for drug-target binding affinity prediction, DeepDTA on the KIBA dataset with a statistical significance. This indicates that the word-based sequence representation adapted by WideDTA is a promising alternative to the character-based sequence representation approach in deep learning models for binding affinity prediction, such as the one used in DeepDTA.  In addition, the results showed that, given the protein sequence and ligand SMILES, the inclusion of protein domain and motif information as well as ligand maximum common substructure words do not provide additional useful information for the deep learning model. Interestingly, however, using only domain and motif information to represent proteins achieved similar performance to using the full protein sequence, suggesting that important binding relevant information is contained within the protein motifs and domains. 
\end{abstract}

% keywords can be removed
% \keywords{First keyword \and Second keyword \and More}

\section{Introduction}

Discovery of potential drugs for new targets is an expensive and time consuming process. Even though over 97M compounds are deposited in PubChem database \cite{bolton2008pubchem} (accession date: Jan 2019), the latest version (version 5.1.2) of DrugBank \cite{wishart2006drugbank} reports only around 12K drug entries. Considering the expansive search space, the development of methodologies to predict the interactions between drugs and targets with high precision can be accelerated by the recent advances in the artificial intelligence applications in chemical research.

Over the last decade, most studies employing traditional machine learning algorithms  modelled the prediction of the interaction between compounds and proteins as a binary classification problem (interacts or not) \cite{yamanishi2010drug,liu2016neighborhood,nascimento2016multiple,keum2017self, greenside2017prediction}. Recently, deep learning architectures have also become a popular choice in drug discovery studies. The success of the first studies \cite{unterthiner2014deep, tian2015boosting}  that employed deep neural networks (DNN) to model the interaction between proteins and compounds  over traditional machine-learning methods motivated later studies that adopted new architectures such as convolutional neural networks (CNNs), recurrent neural networks (RNNs) \cite{ mayr2018large, gao2018interpretable} and  stacked-autoencoders \cite{wang2017computational}. 

Binding affinity value, reported in several different metrics such as disassociation constant ($K_d$), inhibition constant ($K_i$) or  half maximal inhibitory constant ($IC50$), indicates the strength of the interaction between a ligand-protein pair. Prediction of this  valuable information about the interaction became a topic of interest  later, with the studies that  adopted traditional machine learning algorithms such as Kronecker Regularized Least Squares (KronRLS) and boosting machines (SimBoost) \cite{pahikkala2014toward, he2017simboost}.  Deep-learning based studies that aimed to predict bioactivity values of drug-target complexes   utilized DNN \cite{hassan2018dlscore} and  CNN based architectures  using the detailed information that the 3D structures provided \cite{wallach2015atomnet, gomes2017atomic, stepniewska2018development, jimenez2018k}.  
While the information that can be obtained from the 3D structure is very valuable in highlighting the mechanistic information about the bimolecular interaction, these studies depend on the availability of protein - ligand complex structures. The 3D structure  of ligands in complex with a protein is known  for only around 30000 compounds \cite{berman2000protein}.

As an alternative to 3D-based deep learning approaches,  our team proposed a CNN-based model that utilized text-only features of the entities that participate in an interaction (i.e. protein sequences and   the Simplified Molecular Input Line Entry System  (SMILES) notation of ligands) \cite{ozturk2018deepdta} and showed that the binding affinity can be predicted successfully using only raw text data of  protein - compound pairs and  without depending on feature engineering. A deep-learning based model that uses Extended Connectivity Fingerprints (ECFP) and graph-convolutional networks for the representation of compounds, and Protein Sequence Composition (PSC) for protein representation has shown that  ECFP is still a good rival to graph convolutional representation of molecules  \cite{feng2018padme}. This model relies on descriptors that can  be obtained from the SMILES representation.   Another recent protein-compound affinity prediction study showed that compound embeddings obtained with  SMILES-based long-short term memory (LSTM) model performed better than embeddings obtained with graph-convolutional models \cite{mayr2018large}. These findings suggest that a text based approach, which can take advantage of the advances in the natural language processing field, has high potential not only for its simplicity and for the availability of more data, but also because of its representation capability and high performance.

%  \textbf{\cite{karimi2018deepaffinity} This is very similar to what we did but I did not include this because they are using bindingdb data and we couldnt compare the performaces
 
In this study, we propose a methodology to predict protein-ligand binding affinity through text-only information of both proteins and compounds. Without relying on 3D structure information of the complex or 2D  representation of the compound, we learn high dimensional features from sequences of the proteins and ligands. One of the interesting outcomes of the DeepDTA model was   the observation of the difficulty of modelling proteins using their sequences \cite{ozturk2018deepdta}.  When modelled by CNN-based modules separately, the CNN module was not as good at describing proteins as it was with SMILES in the affinity prediction task. We suggested that, since the full length sequence was used, the biologically important short subsequences that would be more powerful at representing the protein were lost due the low signal to noise ratio \cite{ozturk2018deepdta}. In order to overcome this problem, we propose to integrate different pieces of text-based information in the WideDTA model to provide a better representation of the interaction. We still utilize the protein sequence and ligand SMILES string by representing them as a set of   \textit{words}. A word of a protein sequence corresponds to a three-residue subsequence, whereas a word of a ligand is equal to an 8-character subsequence extracted with a sliding window approach \cite{vidal2005lingo}. In addition, we use two textual information sources that can provide valuable clues about the specificity of the interaction. 

The first piece of information we add to our \textit{words} is protein motif and/or domain information. We utilize the PROSITE database \cite{sigrist2009PROSITE} to extract motifs and profiles that are associated with a biologically significant function and domain. Then, we benefit from a recent study that showed that  maximum common substructures (MCS) of ligands constitute the actual  words in the chemical space \cite{wozniak2018linguistic}. Approximately 100K MCS were  used to extract a new set of words from the ligands. Together, these four text-based information sources constitute the WideDTA model.

The results indicate that WideDTA (CI, 0.875) outperforms DeepDTA (CI, 0.863) on the KIBA dataset with statistical significance  (p-value of 0.001) in terms of the CI metric.
 WideDTA, which is built on only the words extracted from protein sequence and ligand SMILES (0.874) also performs better than DeepDTA, which is a character based model, with a statistical significance (p-value of 0.0005) in terms of CI score. We suggest that word-based approach can be a promising alternative to character-based models in this task.

\section{Methods}
\subsection{Dataset}
We used Davis \cite{davis2011comprehensive}, a selectivity assay data for Kinase family proteins, and KIBA \cite{tang2014making} as  benchmark datasets to evaluate the proposed model. The Davis dataset includes the  disassociation constant ($pK_d$) values for about 30K interactions, 69\% of which have affinity values of 10000 nM ($pK_d$=5) indicating weak or no interaction. KIBA, on the other hand, has about three-times more interactions with KIBA scores. KIBA values are computed from the combination of  heterogenous information sources such as $IC_{50}$, $K_i$ and $K_d$. We used the filtered version of the KIBA dataset, in which each protein and ligand has at least ten interactions \cite{he2017simboost}. Table  \ref{Tab:01} summarizes the statistics of the datasets.

\begin{table}[h]
\centering
\caption{Data set } \label{Tab:01} 
\begin{tabular}{@{}llll@{}}
 \hline
Proteins & Compounds & Interactions\\ \hline
Davis ($pK_d$) & 442 & 68 & 30056 \\ \hline
KIBA  & 229 & 2111 & 118254 \\ \hline
\end{tabular}
\end{table}

\subsection{Representation Modules}

In this study, we used four different text-based information sources to model proteins and ligands. 
%Our previous work showed that the use of only protein sequence and ligand SMILES is not adequate  to effectively model the interaction between these entities \cite{ozturk2018deepdta}. Therefore we explored the effect of adding
Our previous work showed that the use of protein sequence and ligand SMILES is an effective approach to model the interactions between these entities \cite{ozturk2018deepdta}. In this study, we explore the effect of adding
additional pieces of specific information, namely domain/motif information for the proteins and maximum common substructure information for the compounds, which might contribute to a better modelling of the interactions. 

\subsubsection{Protein Sequence (PS)}

The protein sequence is composed of 20 different types of amino acids (aa). In this study, we first collected the sequences of the respective proteins from UniProt \cite{apweiler2004uniprot} for each dataset and then, extracted 3-residue words from the sequences similarly to previous studies \cite{asgari2015continuous}.

For instance, an example protein Kinase SGK1 (UniProt, O00141) with the sequence of ``MTVKTEAAKGTLTYSRMRGMVA......YAPPTDSFL" is represented with the following set of  words \{ `MTV', `KTE', ..., `TDS', `TVK', `TEA', ..., `DSF', `VKT', `EAA', ..., `SFL'\}. We will refer to these words that are extracted from the whole protein sequence as PS.

\subsubsection{Protein Domains and Motifs (PDM)}

PROSITE is database that serves as a resource for motif and profile descriptors of the proteins \cite{sigrist2009PROSITE}. Multiple sequence alignment of protein sequences reveals that specific regions within the protein sequence are more conserved than others, and these regions are usually important for folding, binding, catalytic activity or thermodynamics.  These subsequences are called either motifs or profiles. A motif is a short sequence of amino-acids (usually 10-30 aa), while profiles provide a quantitative measure of sequences based on the amino-acids they contain. Profiles are better at detection of domains and families. For instance,  Kinase SGK1 (UniProt, O00141) has the ATP-binding motif `IGKGSFGKVLLARHKAEEVFYAVKVLQKKAILK', while the Protein Kinase Domain profile is about seven times longer than the motif.

We used the PROSITE database to extract motifs and profiles for each respective protein in our datasets. We  then extracted 3-residue subsequences from each motif and domain similarly to the approach adopted in PS. We will refer to this information module as PDM.

% \subsection{%Protein Chemical words (PCW)}

% Protein chemical words are extracted based on a ligand-centric approach that was previously proposed by our team \cite{ozturk2018novel} where a protein is represented with the set of chemical words of its interacting ligands.  In addition, we included a new rule where a protein is represented only with the chemical words of the ligands that they bind to with high affinity (KIBA, value of 12.1; Davis, value of 7.0 as thresholds).  If a protein does not have ligand that it interacts with a high affinity,  then rest of its interacting ligands are considered.}

\subsubsection{Ligand SMILES (LS)}
A chemical compound can be represented with a SMILES string, which is a  syntax that is used to represent atoms, bonds etc. of a molecule with 64 special characters. Canonicalization of a SMILES string is a critical generalization problem because different databases adopt different features in their canonicalization algorithms. Here, we collected the respective SMILES from the PubChem database for all compounds. 

Similarly to protein sequences, we extracted consecutive overlapping k-mers from the SMILES strings. We refer to these k-mers as ``chemical words". For instance the SMILES string ``C(C1CCCCC1)N2CCCC2" is divided into the following 8-character chemical words: ``C(C1CCCC", ``(C1CCCCC", ``C1CCCCC1", ``1CCCCC1), ``CCCCC1)N", ``CCCC1)N2", ... , ``)N2CCCC2". We experimented with $k$ ranging between the values of 7-12 and there was no statistically significant difference in the prediction performance, therefore we chose 8 to be the character length similar to our previous work \cite{ozturk2018novel}. A recent study by \cite{wozniak2018linguistic} also showed that most of the maximum common substructures (MCS) in drugs had between 8 and 12 characters. 

We  utilized a recent methodology that modifies the syntax of SMILES representation, namely DeepSMILES \cite{o2018deepsmiles}. DeepSMILES modifies the use of  parentheses and ring closure digits in the regular SMILES and aims to enhance the performance of the machine learning algorithms that employ SMILES notation as input in various different tasks. We first extracted the canonical SMILES of the compounds from PubChem \cite{wang2016pubchem}, and then converted each SMILES to DeepSMILES (version 1.0.1). For instance the SMILES string ``C(C1CCCCC1)N2CCCC2" is  represented as ``CCCCCCC6))))))NCCCC5" and the chemical words are: ``CCCCCCC6", ``CCCCCC6)", ``CCCCC6))", ... , ``))NCCCC5" with DeepSMILES.

\subsubsection{Ligand Maximum Common Substructures (LMCS)}

\cite{wozniak2018linguistic} recently published an interesting study on the properties of ``words" in chemical space. Rather than using the widely accepted functional groups as chemical words, they speculated that the patterns that the chemists use to distinguish sets of molecules, such as maximum common substructures (MCS), are the ``chemical words" \cite{wozniak2018linguistic}. In order to construct the vocabulary of chemical words, the authors performed a pairwise comparison of approximately 2K molecules. In this study, we used the top 100K most frequent MCS, kindly provided to us by the authors to extract chemical words from the SMILES strings of the ligands in our datasets. 

For instance,  for the example SMILES string  ``C(C1CCCCC1)N2CCCC2", we extract the following MCS \{ `CCCCC', `CCCC', `CCC', `CC', `C1CCCCC1'\}. However, since we represent the SMILES in DeepSMILES syntax, we actually work with  ``CCCCCCC6))))))NCCCC5". Thus, to be able to extract MCS from DeepSMILES syntax, we also converted each MCS into DeepSMILES syntax. Eventually, we extracted the following set of  MCS \{ `CCCCCC', `CCCCCCC', `CCCCC', `CCCC', `CCC', `CC', `CCCCCCC6', `CCCCCC6' \}. The number of MCS captured with DeepSMILES was thus higher, due to the absence of parentheses and ring numbers. 

We will refer to the inputs obtained from MCS as LMCS throughout the article.

\subsection{Representation of text-based modules}
In this study, we proposed a word-based model instead of a character based model because of two reasons: (i)  motifs and domains that were extracted from a protein sequence were not sequential and they can contain overlapping residues, (ii) MCS words can contain overlapping characters. 

\begin{figure*}[h]
\center
\includegraphics[scale=0.5]{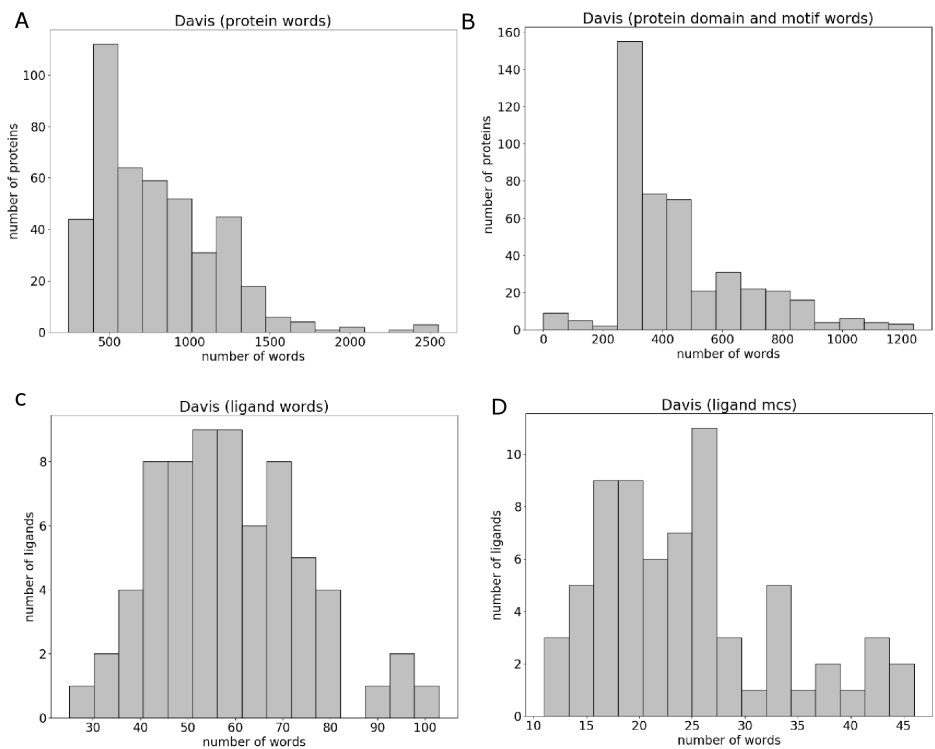}
\caption{Davis dataset - Distribution of number of words.}
\label{fig:wordsdavis}
\end{figure*}

\begin{figure*}[h]
\center
\includegraphics[scale=0.5]{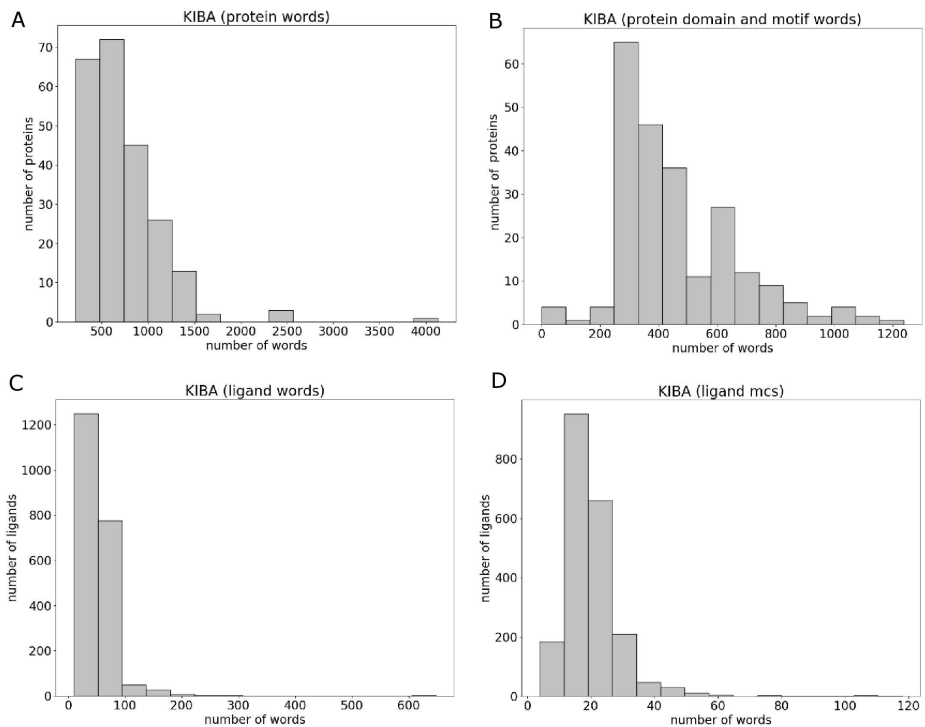}
\caption{KIBA dataset - Distribution of number of words.}
\label{fig:wordskiba}
\end{figure*}

Figures \ref{fig:wordsdavis} and \ref{fig:wordskiba} illustrate the distribution of the number of words that are extracted from each information source versus how many proteins/ligands contain the corresponding number of words for the Davis and KIBA datasets, respectively.

For instance in Figure \ref{fig:wordsdavis}A, we observe that more than 100 proteins have around 500 three-residued words, which also gives us an opinion about the length of an average protein (consisting of words).  We should note that these words are not unique, but might re-occur in the same protein/ligand. We can also articulate that for both of the datasets protein sequences produce the most number of words, whereas MCS produces the least.

As for modeling these words,  each information source was fed into a respective module that consists of two 1D-convolutional layers with a max pooling layer on top, which in turn output abstract features from these raw text inputs.

We used  integer/label encoding, which  uses integers  for the categories to  represent the inputs. We assigned a unique integer to the words that we extracted from the sequences.  For instance, for the LS module our sample SMILES ``C(C1CCCCC1)N2CCCC2" was represented as follows:
 \begin{align*}
   \begin{bmatrix}
           C(C1CCCC  & (C1CCCCC & C1CCCCC1 &  ... & )N2CCCC2 \\
           		1 & 3 & 8  & ... & 71
        \end{bmatrix}
        \end{align*}
in which each integer is associated with a unique 8-character word.

\subsection{Prediction Model: WideDTA}

In this study, we  propose a CNN-based model, which we call WideDTA, that combines at most four pieces of different textual information. We used a similar architecture to the  CNN-based DeepDTA method, which was previously proposed by our team \cite{ozturk2018deepdta} (https://github.com/hkmztrk/DeepDTA). We should note that DeepDTA is a character based model, whereas WideDTA depends on words as input. 

\begin{figure*}[h]
\center
\includegraphics[scale=0.3]{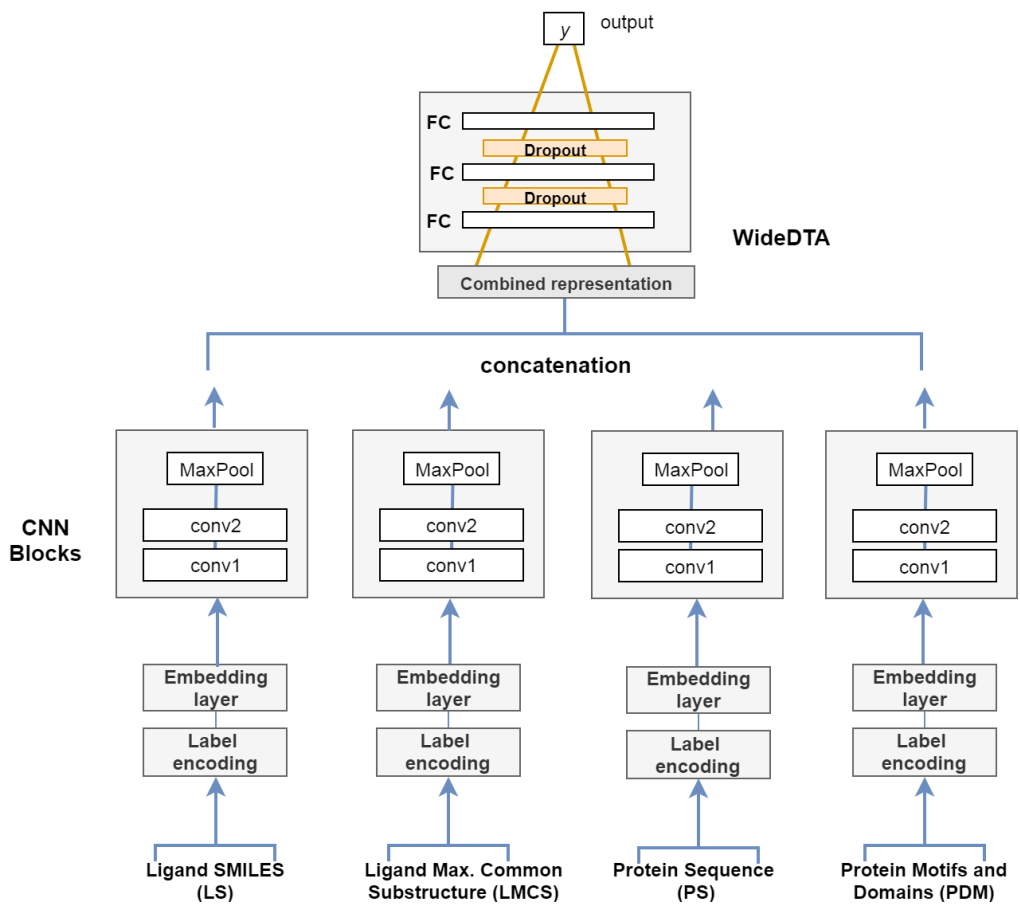}
\caption{Combination of CNN modules constructs the WideDTA architecture.}
\label{fig:02}
\end{figure*}

For each text-based information module, we used two 1D-convolutional layers  with a  max pooling layer on top and Rectified Linear Unit (RELU) as the activation function. We used 32 filters in the first CNN layer, and 64 in the second level CNN in order to capture  more specific patterns. Features extracted from these blocks were concatenated and fed into three fully connected layers with the number of nodes (1024, 1024, 512), which had two drop-out layers in between (value of 0.3) to prevent over-fitting. The proposed model that combines a total of four CNN blocks is illustrated in Figure \ref{fig:02}.

We used  Keras' Embedding layer to represent words with 128-dimensional dense vectors to fed integer encoded inputs. The input for the Davis data set consisted of (85,128),  (1000, 128), (500, 128),   and (30, 128) dimensional matrices for the compounds and proteins, domains and motifs, and MCS, respectively. We represented the KIBA data set with a (100,128), (1000, 128), (650, 128), (35, 128) dimensional matrices for the  compounds, proteins, domains and motifs,  and compound MCS, respectively. We chose the maximum lengths based on the distribution of the words for each dataset,  illustrated in Figures \ref{fig:wordsdavis} and \ref{fig:wordskiba}. 

The model was developed on Keras environment \cite{chollet2015keras} with Tensorflow \cite{abadi2016tensorflow}. We evaluated the performance of the presented models on the benchmark datasets Davis \cite{davis2011comprehensive} and KIBA \cite{tang2014making} and used the same training and test folds for all experiments.  In these folds, both datasets were  randomly divided  into six equal parts and one part was separated as the  independent test set. The remaining parts of the data sets were used to determine the hyper-parameters with cross-validation.

\subsection{Evaluation}
The performance of the proposed model was measured by calculating the Concordance Index (CI)  and Mean Squared Error (MSE) metrics. CI evaluates the ranking performance of  the models that output continuous values  \cite{gonen2005concordance}:
\begin{equation} \label{eq:10}
CI = \frac {1} {Z} \sum_{\delta_x > \delta_y} h (b_x - b_y)
\end{equation}

where $b_x$ is the prediction value for the larger affinity $\delta_x$, $b_y$ is the prediction value for the smaller affinity $\delta_y$, $Z$ is a normalization constant, $h(m)$ is the step function \cite{pahikkala2014toward}:

 $$ 
   h(m) = 
\begin{cases} 
          1,  & \text{if}\ m > 0\\
         0.5,  & \text{if}\ m = 0\\
	   0, & \text{if}\ m <0 
 \end{cases}                                     
$$ 
 
MSE measures the difference between the   predicted values ($p$)  and  the vector of actual values ($y$). For the Davis dataset the compared values are $pK_d$ while for the KIBA dataset the values are $KIBA$ scores. $n$ indicates the number of samples.
\begin{equation} \label{eq:11}
MSE= \dfrac{1}{n}\sum_{k=1}^{n}(p_k - y_k)^2   
\end{equation} 
We also utilized Pearson correlation coefficient, which is explained in Equation \ref{eq:12},  to measure the difference between the true values and the predicted values of binding affinity \cite{kullback1951information}. $cov$ indicates the co-variance between  predicted values $p$ and original values $y$, where $\sigma$ represents the standard deviation.

\begin{equation} \label{eq:12}
Pearson = \dfrac{cov(p, y)   }{\sigma p \sigma y}
\end{equation} 
Statistical significance tests were performed by  paired t-test with  95\% confidence interval.

\subsection{Baseline}

We compared the method presented here with two  state of the art models that employ traditional machine learning  methods.  The first study uses Kronecker-Regularized Least Squares (KronRLS) algorithm to predict binding affinity  in which both proteins and compounds are represented with their pairwise similarity score matrices \cite{pahikkala2014toward}.  In order to compute similarity between proteins and between compounds, Smith-Waterman (S-W) algorithm and  PubChem structure clustering  tool (http://pubchem.ncbi.nlm.nih.gov) were utilized, respectively. In the second study, a gradient boosting machine based method, namely SimBoost, is employed for the prediction of the binding affinity \cite{he2017simboost}. The presented approach depends on feature engineering of compounds and proteins utilizing information such as similarity and network-inferred statistics.  We also compared our results with DeepDTA \cite{ozturk2018deepdta}, which is a deep-learning based method that outperformed the aforementioned methods. 

\section{Results}

In this study, we introduce a model to predict the binding affinity of the interactions between proteins and ligands using a deep learning based model, WideDTA.  The proposed model incorporates up to four text-based information sources: complete sequence,  motif and domain sequences for  proteins, SMILES-based words and maximum common substructures (MCS) for ligands. We should note that DeepSMILES syntax was used throughout the work and SMILES here refers to DeepSMILES.

 Tables \ref{tab:davis1}  and \ref{tab:kiba1} report the average MSE, CI and Pearson scores over the independent test set utilizing different set of features for the prediction of binding affinity for the Davis and KIBA datasets, respectively. We assessed the effect of using different features, starting with different pairs of information sources such as protein sequence and ligand SMILES or protein sequence and MCS, and then adding each feature one by one until the four modules are combined. 

In the Davis dataset, we obtain CI value of 0.875 and MSE of 0.295 when we use full length protein sequence (PS) and ligand SMILES (LS). Comparison of the pairwise models favors two modules, PDM (protein domain and motif) and LMCS (ligand maximum common substructure) with higher CI value (0.883) and lower MSE value (0.276). Even though the improvement over PS (protein sequence) and LS (ligand SMILES) pair is not statistically significant, we choose to append a third information module, PS, to the PDM+LMCS pair because this combination has the best performance. This triplet improves the prediction performance in terms of CI and MSE values.  Finally, combination of the four modules produces the best performance in terms of MSE and Pearson correlation values.
 
 \begin{table*}[h]
 \caption{CI, MSE, and Pearson Correlation Coefficient values for the Davis dataset on the independent test set using the WideDTA model. Standard deviations are given in  parenthesis.} \label{tab:davis1}
 \begin{center}
\scalebox{1}{
\begin{tabular}{|l|l|l|l|l|l|l|l|l|}
\hline
\textbf{}       & \multicolumn{2}{c|}{\textbf{Protein}}     & \multicolumn{2}{c|}{\textbf{Compound}} & \multicolumn{3}{c|}{\textbf{Evaluation}}        \\ \hline
\textbf{Method} & \textbf{PS} &  \textbf{PDM} & \textbf{LS}       & \textbf{LMCS}      & \textbf{CI}   & \textbf{MSE} & \textbf{Pearson} \\ \hline
WideDTA         & 1                     & 0            & 1                 & 0                  & 0.875 (0.008) & 0.295 (0.029) & 0.807 (0.009)     \\ \hline

WideDTA         & 0                    & 1            & 1                 & 0                  & 0.876 (0.009) & 0.311 (0.028) & 0.806 (0.008)     \\ \hline
WideDTA         & 1                      & 0            & 0                 & 1                  & 0.879 (0.003) & 0.284 (0.022) & 0.807 (0.006)     \\ \hline
WideDTA         &   0                & 1             &  0                 &     1           & 0.883 (0.003) &      0.276 (0.009)        &   0.813 (0.003)      \\ \hline
WideDTA         & 1               & 1            & 0                 & 1                  & 0.885 (0.003) & 0.267 (0.002) & 0.814 (0.003)     \\ \hline

WideDTA         & 1                    & 1            & 1                 & 1                  & 0.886 (0.003)  & 0.262 (0.009) & 0.820 (0.006)            \\ \hline
\end{tabular}}
\end{center} 
\end{table*}

As for the KIBA dataset, we observed that PS and LS pair provided the best CI (0.874) and MSE (0.179) values, therefore  this pair was selected as a starting point to add new modules. The performance of the model represented with this pair was higher than the model with PS and LMCS  with a statistical significance in terms of CI,  MSE and Pearson metrics (p-value of  0.0001 for all).  We suggest that LS, which uses chemical words extracted from SMILES, may be a better representation module than LMCS, the maximum common substructure words.  The inclusion of PDM  does not significantly affect the performance, nor does the inclusion of the PDM and LMCS together, even though there are slight improvements with the three-module and four-module systems.

 \begin{table*}[h]
 \caption{CI, MSE, and Pearson Correlation Coefficient values for the KIBA dataset on the independent test set using the WideDTA model. Standard deviations are given in the parenthesis.} \label{tab:kiba1}
 \begin{center}
\scalebox{1}{
\begin{tabular}{|l|l|l|l|l|l|l|l|l|}
\hline
\textbf{}       & \multicolumn{2}{c|}{\textbf{Protein}}     & \multicolumn{2}{c|}{\textbf{Compound}} & \multicolumn{3}{c|}{\textbf{Evaluation}}          \\ \hline
\textbf{Method} & \textbf{PS} &  \textbf{PDM} & \textbf{LS}       & \textbf{LMCS}      & \textbf{CI}    & \textbf{MSE}  & \textbf{Pearson} \\ \hline
WideDTA         & 1                   & 0            & 1                 & 0                  & 0.874 (0.0002) & 0.179 (0.005)  & 0.855 (0.002)            \\ \hline

WideDTA         & 0                  & 1            & 1                 & 0                  & 0.873 (0.001)  & 0.182 (0.004)   & 0.851 (0.002)           \\ \hline
WideDTA         & 1                   & 0            &      0             & 1                  & 0.850 (0.002)  & 0.238 (0.007)  & 0.803 (0.006)            \\ \hline

WideDTA         & 0                   & 1            &      0             & 1                  & 0.849 (0.002)  & 0.237 (0.006)  & 0.805 (0.003)            \\ \hline

WideDTA         & 1                   & 1            & 1                 & 0                  & 0.873 (0.001)  & 0.180 (0.002)  & 0.854 (0.001)            \\ \hline
WideDTA         & 1                    & 1            & 1                 & 1                  & 0.875 (0.001)  & 0.179 (0.008) & 0.856 (0.003)            \\ \hline

\end{tabular}}
\end{center} 
\end{table*}

Inclusion of domain and motif information for proteins did not lead to a significant improvement in prediction ability in either dataset. This may be an expected result for these datasets, both of which belong to the kinase family, whose members have similar motifs and domains. In both datasets, the four major domains and motifs that dominated the information were:  kinase domain (PROSITE ID: PS50011),  kinases ATP-binding region signature (PROSITE ID: PS00107), Serine/Threonine protein kinases active-site signature (PROSITE ID: PS00108) and Tyrosine protein kinases specific active-site signature (PROSITE ID: PS00109). 

We should however highlight that the models that use domains and motifs alone provided similar performances to the models that use full protein sequence. Thus, we can suggest the use of domain and motif information, which in fact constitute a small portion of a sequence, might be a more informative alternative to full sequence. Even though sharing the aforementioned major domains and motifs, majority of the domains and motifs extracted from both datasets (Davis, 70\% and KIBA, 75\%) appear in five or less ligands which is an indication of motifs and domains capturing the distinctiveness.  

We observed that the use of maximum common substructure words provided a slight improvement over SMILES based 8-character words (LS) for the Davis dataset, but led to a significantly worse performance for the KIBA dataset. We might still argue that MCS words distinguish between the characteristics of different molecules considering that the percentage of the unique MCS words that appear in five or less ligands comprise the majority of all MCS words that are extracted from both datasets (Davis, 71\%  and KIBA, 69\% ). However, LS has more of an advantage over MCS words by means of the sliding window approach that might produce words with slight character changes.

We finally  compared the performance of the best WideDTA combination with traditional machine learning based methods KronRLS \cite{pahikkala2014toward} and SimBoost \cite{he2010predict} and deep-learning based state of the art DeepDTA \cite{ozturk2018deepdta}, which is a character-based model that utilizes the complete sequence of proteins and the SMILES of the compounds. Tables \ref{tab:davis2} and \ref{tab:kiba2} report the performance comparison in terms of CI and MSE metrics. 

\begin{table}[h]
\caption{CI and MSE values for Davis dataset on the independent test set for WideDTA and other state-of-art models. Standard deviations are given in the parenthesis.} \label{tab:davis2}
\begin{center}
\scalebox{1.05}{
\begin{tabular}{|c|c|c|c|c|}\hline
Method & Proteins  & Compounds & CI & MSE  \\ \hline
KronRLS  & S-W & PubChem Sim&0.871 (0.0008) & 	0.379 \\ \hline
SimBoost  &  S-W & PubChem Sim & 0.872 (0.002)  &	0.282\\ \hline

DeepDTA  &  PS (char) & LS (char) & 0.878 (0.004)  & 0.261	 \\ \hline

WideDTA (best)  & PS + PDM  & LS + LMCS  & 0.886 (0.003)  & 0.262  	 \\ \hline

\end{tabular}}
\end{center} 
\end{table}

On both datasets, the best combinations of WideDTA outperforms the current state of the art machine learning methods, KronRLS and SimBoost, as well as the deep learning based approach DeepDTA.  On the KIBA dataset, the best model produces a CI score (0.875)  better than DeepDTA (0.863) with statistical significance (p-value of 0.001). WideDTA with only LS and PS as information  (CI score of 0.874) still outperforms DeepDTA (CI score 0.863), which is a character based model with a statistical significance (p-value of 0.0005).  We observe that use of words instead of characters can be more informative for the task of protein-ligand interaction prediction.

\begin{table}[h]
\caption{CI and MSE values for KIBA dataset on the independent test set for WideDTA and other state of the art models. Standard deviations are given in the parenthesis. } \label{tab:kiba2}
\begin{center}
\scalebox{1.05}{
\begin{tabular}{|c|c|c|c|c|}\hline
Method & Proteins & Compounds &  CI & MSE\\ \hline
KronRLS 	 &   S-W &	Pubchem Sim &0.782 (0.0009) &	0.411 \\ \hline
SimBoost 	 & S-W  &  Pubchem Sim&	0.836 (0.001) &	0.222 \\ \hline

DeepDTA  &  PS (char) &  LS (char) & 0.863 (0.002)  & 0.194	 \\ \hline

WideDTA (best)  & PS + PDM  & LS + LMCS  & 0.875 (0.001)  & 0.179   \\ \hline

\end{tabular}}
\end{center}
\end{table}

%\subsubsection{%SMILES Enumeration}

%Following,  we will integrate enumerated SMILES to training data in order to observe the affect of the training size to the prediction performance.

%\begin{table}[H]
%\caption{Effect of the length of chemical words (cw) extracted from SMILES on Davis dataset with CI and MSE values as performance metrics. Standard deviations are given in the parenthesis.} \label{tab:davis2}
%\begin{center}
%\scalebox{1.05}{
%\begin{tabular}{|c|c|c|c|c|}\hline
%Method &  Features & cw length & CI & MSE  \\ \hline
%WideDTA  & PS + LS   &  x1 & 0.868 (0.007)  & 0.306 (0.x)	 \\ \hline
%WideDTA  & PS + LS     &  x3 & 0.877 (0.004)  & 0.287 (0.x)	 \\ \hline
%WideDTA  & PS + LS  & x5 &  0.873 (0.006) & 0.263 (0.x)	 \\ \hline
%\end{tabular}}
%\end{center} 
%\end{table}

\subsection{Biological Interpretation}

% Table \ref{tab:protbased} reports the performance of the best models of the WideDTA for each dataset in which the CI and MSE values are computed and then averaged for each protein. This metric gives us information about the predictive power of the methodology for a single protein and all its interacting ligands. We can observe the distinct decrease in the performance, more particularly in CI metric. Since concordance index (CI) is a ranking-oriented evaluation metric, when considering ranking of a rather small range of pairs (each drug interaction of a protein), it is expected that the performance might drop.

%\begin{table}[H]
%\caption{ Protein-based average CI and MSE values for best performing combinations of Davis and KIBA datasets . Standard deviations are given in the parenthesis.} \label{tab:protbased}
%\begin{center}
%\scalebox{1.05}{
%\begin{tabular}{|c|c|c|c|c|}\hline
%Dataset & Method &  Features  & CI & MSE  \\ \hline
%Davis & WideDTA  & PS + PDM + LS + LMCS   &  0.810 & 0.267	 \\ \hline
%KIBA & WideDTA  & PS + PDM + LS + LMC  & 0.809   & 0.296	 \\ \hline

%\end{tabular}}
%\end{center} 
%\end{table}

We investigated a sample protein, human cell cycle checkpoint kinase Chk1 (UniProt, O14757), from KIBA dataset and its corresponding predictions by comparing the performance of the model where all modules are included with the model where only PDM and LMCS modules are used in order to elucidate why MCS words fail to describe the molecules in the KIBA dataset.

We selected a case where PS + PDM + LS + LMCS achieved better performance (predicted, $10.60, 10.47$) at predicting the actual weak  binding affinities (real value, $10.20, 10.40$) for two sample ligands compared to the PDM + LMCS model (predicted, $11.34, 11.18$).

%We selected a case for two ligandswhere PDM + LMCS (predicted, $11.34, 11.18$) failed to predict actual weak  binding affinities (real value, $10.20, 10.40$) accurately whereas the use of PS + PDM + LS + LMCS (predicted, $10.60, 10.47$) succeeded. 
We inspected these two ligands, Pubchem IDs 404051 (N-(3-Chlorophenyl)-4-methoxybenzamide) and 324081 (4-Anilinoquinazoline) in terms of chemical words and MCS words. Even though MCS words succeeded in capturing the following pattern in two of the ligands, ``C=CC=CC=C", LS chemical words were also able to capture the similar pattern with ``C=CC=CC=C6))", but  with several words such as  `C=CC=CC=', `=CC=CC=C' and `CC=CC=C6'. However LS also uncovered  words such as `NC=NC=NC' and `NC=NC=CC'  that point out a similar substructure. Since these chemical words are created with a sliding window approach, every character change (e.g. an atom at the end of the word) leads to a new word. Therefore, we  suggest that comprising words that are one character away might help filters of the CNN to capture the similarity between the words such as `NC=NC=NC' and `NC=NC=CC'.

\section{Conclusion}

With this study, we proposed a deep-learning based approach to predict drug-target binding affinity, which we refer to as WideDTA. WideDTA combines four different textual pieces of information related to proteins and ligands. For proteins we used the complete amino acid sequences as well as the domains and motifs extracted from the PROSITE database. The protein sequence, domains and motifs were all represented as a set of 3-residue words, whereas for ligands, 8-character chemical words and maximum common substructure (MCS) words were extracted from the complete SMILES strings. 

The results showed that even in the absence of information provided by domains and motifs and MCS words, protein sequence and ligand SMILES text alone provided comparable and significantly better performances than the state of the art deep learning approach, namely DeepDTA, on the Davis and KIBA datasets. DeepDTA is a character based approach \cite{ozturk2018deepdta}, whereas WideDTA uses word representations. Our results suggest that the word based approach has higher performance than a character based approach.

%% TODO: using domains and motifs were as well as using whole protein sequence
Despite our expectation that adding the protein domain and motif information extracted from the PROSITE database would improve the protein representation and hence the prediction performance, we could not observe a statistically significant improvement in the predictive power with the addition of this information. We realize that one limitation of the current study is its focus on using two kinase benchmark datasets. Due to the similarity in the kinase structure and motifs, including domain and motif information to the full protein sequence information does not provide additional information that would distinguish the different proteins in the dataset. However, this information can be informative in datasets that are more diverse. %Our future work is to test WideDTA on a dataset collected from databases such as ChEMBL or BindingDB \cite{liu2007bindingdb}.

We should however emphasize an interesting outcome of this work. The use of protein domains and motifs only, which correspond to a  smaller percentage of the full sequence, performed as well as the use of the complete protein sequence. This is also supported by the dataset statistics that indicated that the majority of the unique domains and motifs extracted from the proteins appear in  five or less ligands, which in turn points out the captured individuality of the proteins.

\section*{Acknowledgements}
TUBITAK-BIDEB 2211-E Scholarship Program (to H.O.) and BAGEP Award of the Science Academy (to A.O.) are gratefully acknowledged.  We thank Maciej Eder and Nano Grzybowski for sharing the MCS vocabulary. 

\section*{Funding}
This work was supported by Bogazici University Research Fund (BAP) Grant Number 12304.

\bibliographystyle{unsrt}  
\bibliography{references}  %%% Remove comment to use the 
\end{document}